
\def\\{\hfil\break}

\tolerance=8000
\hsize=15.2truecm
\parskip=4pt

\def\newpage{\vfill\eject}

\def\\{\hfil\break}

\hsize=15.2truecm
\parskip=4pt

\def\newpage{\vfill\eject}

\def\\{\hfil\break}

\vskip 3 cm

\noindent

\line{\hfil			                  HU-TFT-94-01}
\line{\hfil                                     3 January 1994}
\vskip 3.0truecm

\centerline{\bf  A HIERARCHY OF EFFECTIVE FIELD THEORIES}
\centerline{\bf OF HOT ELECTROWEAK MATTER}
\bigskip\bigskip\bigskip
\centerline{ A. Jakov\'ac$ ^{a}$, K. Kajantie$ ^{b}$ and A. Patk\'os$ ^{a,c}
$ }
\medskip
\centerline {\sl $ ^{a}$ Dept. of Atomic Physics, E\"otv\"os Univ., Budapest}
\centerline {\sl $ ^{b}$Dept. of Theor. Phys., University of Helsinki}
\centerline {\sl $ ^{c}$The Niels Bohr Institute, University of Copenhagen}
\medskip

\vfill

{\narrower\noindent {\bf Abst\-ract:} A hierarchy of effective
three-dimensional theories of finite temperature electroweak matter
is studied. First an integration over non-static modes
leads to an effective theory containing a gauge field
$A_{i}^{a}$, an adjoint Higgs field $A_{0}^a$
and the fundamental Higgs field $\phi_{\alpha}$. We carry out the
integration in the 1-loop approximation, study renormalisation effects
and estimate quantitatively those
terms of the effective action which are suppressed by inverse
powers of the temperature. Secondly, because of the existence of
well-separated thermal mass scales, $A_{0}^a$ can be integrated over,
and finally also $\phi_{\alpha}$, leaving an effective theory of the
$A_{i}^{a}$.
In the analysis of the subsequent models particular
attention is paid to the screening of the magnetic fluctuations due to the
integrated-out degrees of freedom.}
\vfill
\hrule
\smallskip\noindent
E-mail: jako@hercules.elte.hu, kajantie@phcu.helsinki.fi,
patkos@ludens.elte.hu
\newpage
\bigskip\bigskip
{\bf 1. Introduction}
\bigskip
The restoration of the global gauge symmetry of electroweak interactions
appears to be a particularly weak first order phenomenon[1-5].
The general experience with phase transition phenomena gives the intuition that
the lightest thermal fluctuations (of the longest correlation length)
play decisive role in the transition range. Renormalisation group
techniques suggest that the role of the more "massive" fluctuations
is to influence the effective dynamics of the light modes.
In the standard model of electroweak interactions one encounters several mass
scales, all varying with temperature. Therefore the search for the
existence of scale hierarchies should be realised in the process of
solving the model.

{}From this point of view the elimination of the non-zero Matsubara
frequency modes [6,7] represents only a first step of the gradual reduction.
It rearranges the usual perturbative series of the effective potential.
For instance, even if the integration is performed in the Gaussian
approximation, it brings into the tree-level expression of the effective
three-dimensional potential perturbatively higher order contributions.
In section 2 of the present paper we are going to discuss in great
detail the importance of different terms of the effective Lagrangian,
arising from the
expansion of the 1-loop expression of the contribution from the non-static
modes.

The main tool
of this investigation is the comparison of the characteristics of the
electroweak transition computed perturbatively within different versions
 of the effective 3D theory. The leading result of two subsequent
1-loop integrations
is known to be equivalent to the summation of the leading infrared sensitive
loop contributions in the 4D thermal theory. Subleading effects related
to taking into account higher
dimensional field combinations in the effective 3D theory and/or to the
implementation of specific 4D renormalisation conditions for the
$T$-independent part of the potential energy density, are discussed for
different ranges of the Higgs-mass. Efforts will be made to clarify relevant
technical details of the evaluation and renormalisation of the 1-loop
fluctuation determinant.

The general conclusion of this part of our study is that the subleading
 corrections
have increasingly important  impact on the transition as the Higgs mass
is lowered. However, for $m_{H}> 50$ GeV one can
safely omit them from the the effective model.

The perturbative solution of the 3D model suggests that distinct mass-scales
exist also among the static fluctuation modes.
In section 3 further thinning
of the degrees of freedom will be proposed on the basis of this observation.
The first field to be eliminated is the adjoint Higgs field
$A_0^a$. It gets screened already through the
integration of the non-static modes, and is found to be much more massive
 than other fluctuations for any Higgs mass.

In the experimentally relevant range $ 60$ GeV $\leq m_{H}$ the existence of
further distinct scales is not obvious. Only above 80 GeV
can one convincingly argue that
the perturbatively determined thermal masses of the scalar
$\phi$ fields are 3-4 times larger near
$T_{c}$ than $m_{W}(T=T_c)$. Then the integration over these scalar fields
is also justified.
The main results of this part of our paper are the
propositions for effective 3D gauge field-Higgs and pure gauge theories
(Eqs. (29) and (34)), respectively.

In the course of the
gradual integrations one can observe many pieces of the infrared
improved self-energies, familiar from the corresponding 1-loop Schwinger-Dyson
equations of the full thermal theory [3,4], to emerge. Moreover,
the non-zero scalar
background breaking the global gauge symmetry, induces not just a piece of
mass
for the magnetic vector potential but also higher "$A^{2n}$" potential terms do
appear. It is remarkable that working with non-Abelian constant vector
background the finite $T$ rescaling of the gauge kinetic term ("wave-function
renormalisation") can be obtained, too.

 Non-perturbative methods are being applied to the study of the 3D effective
model of the electroweak transition [8,9], with indications for
important non-per\-tur\-ba\-ti\-ve effects. The effective 3D
gauge field-Higgs and pure gauge field models
resulting from succesively integrating out $A_{0}^a$ and $\phi$ would merit
non-perturbative studies of their own.

Section 4 is devoted to the discussion of the description of the
electroweak transition emerging from the perturbative solution of the effective
pure gauge theory. In spite of the crudeness of the approximation  it
predicts stronger discontinuities at the phase transition in comparison with
the perturbative solution of the
full finite $T$ theory. This feature might hint to the true characteristics
of the more complete description.

On the basis of the present work the region of small ($\leq 30$ GeV)
Higgs masses cannot be described adequately with a pure gauge effective model.
We found evidence that the appropriate theory of the system in this regime
is rather an effective scalar model.
 The investigation of this theory with help of non-perturbative
numerical methods is the subject of a parallel project [11].
\bigskip
{\bf 2. Virtues of the effective theory of static modes}
\bigskip
The decoupling of non-zero frequency (non-static) Matsubara modes from
the high temperature thermodynamics of field theories has been proposed
long time ago [6] and has been studied in detail, for example, in [7].

It is well-known, that the leading thermal (screening) masses are
correctly induced by integrating over the non-static modes at 1-loop.
It is, however, less widely appreciated that the 1-loop Schwinger-Dyson
equations for the self-energies of the different fields, as calculated
from the effective 3D theory coincide with the 1-loop equations of the
full thermal theory. This means, that the infrared improvement program
for the perturbation theory of the effective Higgs-potential
[3,4] can be implemented in the effective model, with equivalent results.

{}From our point of view the role of the integration over the non-static
modes is not only to improve the infrared behavior of the propagators,
but also to induce new, highly non-linear interactions between the static
fields. These arise when the logarithm of the determinant of the 1-loop
functional fluctuations is expanded into power series with respect to
the background fields. At $T=\infty$ all combinations with
dimensions higher than 4 are negligible.
 The appearence of these terms (formally of order $g^{6}, \lambda^{3}$)
in the tree-level 3D action
illustrates how the gradual integration reorganizes the perturbative
series. At the realistic electroweak phase
transition the importance of these terms should be quantitatively
assessed. An attempt for this in the U(1) and SU(2) Higgs models is the
content of this section.

The models under consideration are the following:

G=SU(2):
$$S=\int_{0}^{\beta}\int d^{3}x\bigl [{1\over 4}F_{mn}^{a}F_{mn}^{a}+
{1\over 2}(D_{m}\Phi)^{+}(D_{m}\Phi)+{1\over 2}m^{2}\Phi^{+}\Phi\
+{\lambda\over 24}(\Phi^{+}\Phi)^{2}\bigr ]$$
$$~~~~~~~~~~~~~~~~~~~~~~~~~+{\rm counter~terms},\eqno(1)$$

G=U(1):
$$S=\int_{0}^{\beta}\int d^{3}x \bigl[{1\over 4}F_{mn}F_{mn}+
{1\over 2}|(\partial_{m}+igA_{m})\Phi|^{2}+{1\over 2}m^{2}|\Phi |^{2}
+{\lambda\over 24}|\Phi |^{4}\bigr]+{\rm counter~terms}\eqno(2)$$
(m=1,..,4; a=1,2,3; $D_{m}\Phi =(\partial_{m}+igA_{m}^{a}\tau^{a}/2)\Phi$).

For definiteness the integration over the non-static fields
 will be discussed in considerable detail for the G=SU(2) case,
the analogous formulae for G=U(1) will be just listed next to it.

The elimination of the non-static modes is most conveniently performed
in the thermal static gauge, which preserves invariance under
space-dependent gauge transformations:
$$A_{0}^{a}(x,\tau )=A_{0}^{a}(x)+a_{0}^{a}(x,\tau ),~~a_{0}^{a}(x,\tau )=
\sum_{n\neq 0}a_{0,n}^{a}(x)e^{i\omega_n\tau},~\omega_{n}=2\pi Tn,$$
$$a_{0}^{a}(x,\tau )\equiv 0,~~A_{0}^{a}(x)\equiv\delta^{a3}A_{0}(x).\eqno(3)$$
Then, it is sufficient to keep in
the computations of the effective action only the static background fields
$A_{0}(x),~\Phi_{0}(x)$, since the dependence on static $A_{i}(x)$ is
easily established with help of the minimal coupling principle.
Furthermore, we are going to investigate only the potential part of the
induced action, therefore in the course of the actual calculations the
background fields will be treated as constants.

Performing the expansion of (1) in the non-static fields up to quadratic terms,
the fluctuations, coupled on the Gaussian level are separated most conveniently
by introducing the spatially transverse $(a_{i,T})$ and longitudinal
($a_{L}\equiv \partial_{i}a_{i}$) components of the vector fluctuations.
Then the groups fluctuating independently in the gauge (3) are
 $(a_{L}^{1},a_{L}^{2},\xi_{1},\xi_{2})$;
 $(a_{L}^{3},\eta_{1},\eta_{2})$;
$(a_{i,T}^{1},a_{i,T}^{2})$;
 $a_{i,T}^{3}$,
 where the fluctuations of the Higgs field around
the background are parametrized as
$$\Phi=(\xi_{1}+i\xi_{2}, \Phi_{0}+\eta_{1}+i\eta_{2}).$$
The matrices of the quadratic forms for the fluctuations are the following,
respectively:

$$
\eqalign{
&K_1=\left(\matrix{
m_L^2&2igA_{0}\omega_n&0&{ig\over2}\Phi_0 k\cr
-2ieA_{0}\omega_n&m_L^2&{ig\over2}\Phi_0 k&0\cr
0&-{ig\over2}\Phi_0 k&M^2&-igA_{0}\omega_n\cr
-{ig\over2}\Phi_0 k&0&igA_{0}\omega_n&M^2\cr
}\right),\cr
&K_2=\left(\matrix{
m_L^{\prime2}&0&-{ig\over2}\Phi_0 k\cr
0&M^{\prime2}&igA_{0}\omega_n\cr
{ig\over2}\Phi_0 k&-igA_{0}\omega_n&M^2\cr
}\right),\cr
&K_3=\left(\matrix{
m_T^2&2igA_{0}\omega_n\cr
-2igA_{0}\omega_n&m_T^2\cr
}\right),\cr
&K_4=\left(\matrix{m_T^{\prime2}\cr}\right),\cr
}\eqno(6)$$
where we use the temporary abbreviations
$$\eqalign{
&m_L^2=\omega_n^2+g^2\left(A_{0}^2+{\Phi_0^2\over4}\right),\cr
&m_L^{\prime2}=\omega_n^2+g^2{\Phi_0^2\over4},\cr
&m_T^2=k^2+\omega_n^2+g^2\left(A_{0}^2+{\Phi_0^2\over4}\right),\cr
&m_T^{\prime2}=k^2+\omega_n^2+g^2{\Phi_0^2\over4},\cr
&M^2=k^2+\omega_n^2+m^2+g^2{A_{0}^2\over4}+\lambda{\Phi_0^2\over6},\cr
&M^{\prime2}=k^2+\omega_n^2+m^2+g^2{A_{0}^2\over4}+\lambda{\Phi_0^2\over2}.\cr
}\eqno(7)$$

After evaluating the four functional determinants,
the expression of the 1-loop non-static contribution to the
non-derivative part of the 3D action can be given as follows:
$$U(A_0,\Phi)=\sum_{n\neq 0}\sum_{k}\bigl\{\ln (K^{2}+{1\over
4}g^{2}\Phi_0^{2})+\ln\bigl [(K^{2}+g^{2}(A_{0}^{2}+{1\over
4}\Phi_0^{2}))^{2}-4g^{2}A_{0}^{2}\omega_{n}^{2}\bigr ]$$
$$+{1\over2}\ln\bigl [(\omega_{n}^{2}+{1\over
4}g^{2}\Phi_0^{2})\bigl ((K^{2}+{1\over 4}g^{2}A_{0}^{2}+{1\over
6}\lambda\Phi_0^{2}+m^{2})(K^{2}+{1\over 4}g^{2}A_{0}^{2}+{1\over 2}\lambda
\Phi_0^{2}+m^{2})$$
$$~~~~~~~~~~-g^{2}A_{0}^{2}\omega_{n}^{2}\bigr)-{1\over
4}g^{2}\Phi_0^{2}k^{2}(K^{2}+{1\over 4}g^{2}A_{0}^{2}+{1\over
2}\lambda\Phi_0^{2}+m^{2})\bigr ]$$
$$~~~~+{1\over 2}\ln\bigl [ (\omega_{n}^{2}+g^{2}(A_{0}^{2}+{1\over
4}\Phi_0^{2}))^{2}\bigl ((K^{2}+{1\over 4}g^{2}A_{0}^{2}+{1\over
6}\lambda\Phi_0^{2}+m^{2})^{2}-g^{2}\omega_{n}^{2}A_{0}^{2}\bigr
)$$
$$~~~~~~~~~~~~-{1\over
2}g^{2}\Phi_0^{2}k^{2}(\omega_{n}^{2}+g^{2}(A_{0}^{2}+{1\over
4}\Phi_0^{2}))(K^{2}+{1\over 4}g^{2}A_{0}^{2}+{1\over
6}\lambda\Phi_0^{2}+m^{2})+4g^{4}A_{0}^{4}\omega^{4}$$
$$~~~~~~~~~~~-4g^{2}A_{0}^{2}\omega_{n}^{2}(K^{2}+{1\over
4}g^{2}A_{0}^{2}+{1\over
6}\lambda\Phi_0^{2}+m^{2})^{2}-g^{4}A_{0}^{2}\Phi_0^{2}
\omega_{n}^{2}k^{2}+{1\over
16}g^{4}\Phi_0^{2}k^{4}\bigr ]\bigr\}.\eqno(8)$$
In this equation $K^{2}=k^{2}+\omega_{n}^{2}.$

In order to find the couplings for the various pieces of the effective
theory one expands the logarithms in polynomials of the background fields.
The expansion is fully justified, since the $n=0$ modes
are left out from the calculation of the sums.
This rather cumbersome calculation is outlined for the more transparent
case $A_{0}=0$ in the Appendix. This case is actually sufficient to
treat the renormalisation of the action.
For this we use the
analogue of Linde's renormalisation conditions [10,12]:
$${dU(T-{\rm indep})\over d\Phi_0}=0,~~{d^{2}U({\rm T-indep})\over
d^{2}\Phi_0}=
m_{H}^{2}(T=0),~~~\phi=v_{0}.\eqno(9)$$
Here $v_{0}$ is the classical vacuum expectation value of the Higgs field
and $m_H(T=0)$ is the physical Higgs mass.

The cut-off regularised expression of the scalar potential is given in
eq.(A.7).
For renormalisation its $T$-dependent part should be separated.
This is achieved by rewriting the term depending logarithmically on
$\Lambda^{2}
/T^{2}$ in the following convenient form:
$$-{1\over 64\pi^{2}}\ln{\Lambda^{2}\over T^{2}}\bigl [ m^{2}\Phi_{0}^{2}
({9g^{2}\over 2}+2\lambda)+\Phi_{0}^{4}({9g^{4}\over 16}+{3g^{2}\lambda\over
4}+{\lambda^{2}\over 3})\bigr ]$$
$$=-{1\over 64\pi^{2}}\sum_{q}n_{q}m_{q}^{4}(\Phi_{0})\bigl [\ln{\Lambda^{2}
\over m_{q}^{2}(v_{0})}+\ln{m_{q}^{2}(v_{0})\over m_{q}^{2}(\Phi_0)}
+\ln{m_{q}^{2}(\Phi_{0})\over T^{2}}\bigr ].\eqno(10)$$
Here $q$ runs through the fields of the theory: $q=T$ (transversal), $L$
(longitudinal), $H$ (Higgs), $G$ (Goldstone), the $m_{q}^{2}(\Phi_{0})$ are
appropriately chosen mass-squared expressions given below and the
$n_{q}$ are the corresponding
multiplicities.

Choosing for the physical fields (the transversal gauge and the Higgs) the
familiar mass expressions (coinciding with the $T=0$ formulae when
$\Phi_{0}=v_{0}$ is substituted)
$$n_{T}=6,~~~~~m_{T}^{2}(\Phi_0)={1\over 4}g^{2}\Phi_{0}^{2},$$
$$n_{H}=1,~~~~~m_{H}^{2}(\Phi_0)=m^{2}+{\lambda\over 2}\Phi_{0}^{2},\eqno(11)$$
then the transcription (10) requires the unique further choice
$$n_{L}=-3,~~~~m_{L}^{2}(\Phi_{0})={1\over \sqrt{2}}g^{2}\Phi_{0}^{2},$$
$$n_G=3,~~~~m_{G}^{2}(\Phi_{0})=m^{2}+({3\over 4}g^{2}+{\lambda\over 6})
\Phi_{0}^{2}.\eqno(12)$$
(Recall from the Appendix, that the fluctuations in these last degrees of
freedom are not eigendirections of the fluctuation matrix, therefore these
"masses" have no direct physical meaning, not even for $T=0$.)

The last term in the square bracket of  the right hand side of (10)
($\sim \ln (m_{q}^{2}(\Phi_{0})/T^{2})$) will be
associated with the temperature dependent part of the effective potential.
The first term clearly is to be absorbed into the counterterms, while the
second gives a finite contribution. Since the renormalisation conditions (9)
are
fulfilled by the non-logarithmic terms separately, also the first two terms
of the expansion of this finite term around $\Phi_{0}=v_{0}$ should be
cancelled by the countertems. Therefore the $T$-independent part of the
potential fulfilling the renormalisation conditions reads as
$$
U_{\rm eff}({\rm T-indep})={1\over 2}m^{2}\Phi^{+}\Phi+{1\over
4}(\Phi^{+}\Phi)^{2}$$
$$~~~~~~~+{1\over 64\pi^{2}}\sum_{q}n_{q}\bigl [m_{q}^{4}(\Phi)(
\ln{m_{q}^{2}(\Phi )\over m_{q}^{2}(v_{0})}-{3\over 2})+2m_{q}^{2}(\Phi)m_{q}^
{2}(v_{0})\bigr ].\eqno(13)$$
Upon adding to it the $T$-dependent part from eq.~(10), the term
$\ln m_{q}^{2}(\Phi)/m_{q}^{2}
(v_{0})$ of
(13) will be replaced by $\ln T^{2}/m_{q}^{2}(v_{0})$.

The final expression
for the effective action with all abbreviations written out  explicitly is
given in the following eqs.(14-17):
$$
S[\bar A_i,\bar A_0, \varphi] = \int d^3x \bigl[L_{\rm kin}+U(\bar A_0,\varphi)
\bigr],
$$
where
$$
L_{\rm kin}={1\over 4}\bar F_{ij}^{a}\bar F_{ij}^{a}+{1\over
2}[(\partial_{i}+i\bar g\bar A_{i})\varphi ]^{+}[(\partial +i\bar
g\bar A_{i})\varphi ]+{1\over 2}(\partial_{i}\bar A_{0}+\bar
g\epsilon^{abc}\bar A_{i}^{b}\bar A_{0}^{c})^{2},
$$
$$
U_{{\rm dim} 4}(\bar A_0,\varphi)={1\over 2}m_{\varphi}^{2}\varphi^{+}\varphi+
{1\over
2}m_{D}^{2}\bar A_{0}^{2}+{1\over 8}\bar
g^{2}\bar A_{0}^{2}\varphi^{+}\varphi+{T\hat\lambda\over
24}(\varphi^{+}\varphi)^{2}$$
$$~~~~~~~~~~+{17\beta\bar g^{4}\over
192\pi^{2}}(\bar A_{0}^{2})^{2}+ {\rm 3D~counter terms},\eqno(14)$$
$$m_{\varphi}^{2}=\hat m^{2}+({3\over 16}\bar g^{2}+{\bar\lambda\over
12})T,~~~~m_{D}^{2}={5\over 6}\bar g^{2}T+{m_{R}^{2}\bar g^{2}\over
8\pi^{2}T},\eqno(15)$$
$$\hat m^{2}=m^{2}
(1-{1\over 32\pi^{2}}\bigl(({9\over 2}g^{2}+\lambda)\ln{3g^{2}v_{0}^{2}
\over 4T^{2}}+\lambda\ln{\lambda v_{0}^{2}\over  3T^{2}}\bigr)
-{1\over 128\pi^{2}}(45g^{2}+20\lambda+{27g^{4}\over \lambda})),$$
$$\hat\lambda =\lambda -{3\over 8\pi^{2}}(g^{4}{3\over 8}(\ln {g^{2}v_{0}^{2}
\over 4T^{2}}-{3\over 2}\ln{g^{2}v_{0}^{2}\over \sqrt{2}T^{2}})+
{\lambda^{2}\over 4}\ln{\lambda v_{0}^{2}\over 3T^{2}}+3({3g^{2}\over 4}+{
\lambda\over 6})^{2}\ln{3g^{2}v_{0}^{2}\over 4T^{2}})-$$
$$~~~~~~~~~~~~~~~~~~-{9\over 16\pi^{2}}({9g^{4}\over 16}+{3g^{2}\lambda\over
4}+{\lambda^{2}\over 3}).\eqno(16)$$

In eqs.(14-16) we have introduced rescaled 3D fields and couplings, with
dimensionalities appropriate to a 3D theory:
$$\varphi=\sqrt{\beta}\Phi,~~\bar A_{i}=\sqrt{\beta}A_{i},~~\bar A_{0}=
\sqrt{\beta}A_{0},~~\bar g^{2}=g^{2}T,~~\bar\lambda =\lambda T.
\eqno(17)$$
The 3D counterterms, fully specified by the reduction step, will cancel
linearly diverging radiative mass contributions to the $\varphi$ and $\bar
A_{0}$ fields of the effective 3D theory. The form (14) of the effective
theory with
only the leading ${\cal O}(T^{2})$ mass corrections and unmodified $\lambda$
has been given in [8].

The magnitude of the logarithmic
corrections to $m^{2}$ and $\lambda$ is illustrated
by Fig. 1, where in the $m_H - T$-plane the contour lines corresponding to
3\%, 5\% and 7\% variations are displayed. The critical region for the
interesting range
of Higgs masses is covered by these lines. The  effect of renormalisation
conditions on the physical charecteristics of the transition will be
discussed further at the end of this section.

The corrections of relative order $m^{2}/T^{2}$ and $g^{2}\Phi^{2}/T^{2}$
can be calculated by retaining one more order in the expansions of the
logarithms of (8). The corresponding dim 6 combinations of $m^{2},
A_{0}^{2}$ and $\Phi^{2}$ provide two types of contributions to the
effective 3D theory. Firstly,
${\cal O}(m^{2}/T^{2})$ corrections will be generated to the thermal masses of
(15) as well as to the couplings $\bar\lambda$ and $\bar g$.
Secondly, new interactions of the
general form
$$U_{{\rm dim}\,6}(\bar A_0,\varphi)={1\over T^{3}}[
a(\varphi^{+}\varphi)^{3}+b(\varphi^{+}\varphi)^{2}\bar A_{0}^{2}+
c\varphi^{+}\varphi
(\bar A_{0}^{2})^{2}+d(\bar A_{0}^{2})^{3}].\eqno(18)
$$
are also produced.
After performing the necessary algebraic manipulations with help of
symbolic program packages one finds the following results for these
new contributions:
$$\delta m_{\phi}^{2}={m^{4}\zeta (3)\over 128T^{3}\pi^{4}}(-{9\over 4}\bar
g^{2}+\bar\lambda),$$
$$\delta m_{D}^{2}=-{m^{4}\zeta(3)\bar g^{2}\over 64T^{3}\pi^{4}},$$
$$\delta\bar\lambda ={m^{2}\zeta (3)\over 32T^{3}\pi^{4}}(-{27\over
16}\bar g^{4}-{9\bar g^{2}\bar\lambda\over 4}+\bar \lambda^{2}),$$
$$\delta\bar g^{2}=-{m^{2}\zeta (3)\bar g^{2}\over 8T^{3}\pi^{4}}(3\bar
g^{2}+{1\over 4}\bar\lambda),$$
$$a={\zeta (3)\over 1024\pi^{4}}({3\over 16}\bar g^{6}-{3\over
8}\bar\lambda\bar g^{4}-{1\over 4}\bar\lambda^{2}\bar g^{2}+{5\over
27}\bar\lambda^{3}),$$
$$b=-{\zeta (3)\bar g^{2}\over 1024\pi^{4}}({109\over 16}\bar
g^{4}+{47\over 6}\bar\lambda\bar g^{2}+{5\over 9}\bar\lambda^{2}),$$
$$c=-{\zeta (3)\bar g^{6}\over 64\pi^{4}},~~~~d=0.\eqno(19)$$
(Here $\delta\bar\lambda$ and $\delta\bar g^{2}$ refer to the finite
corrections of the coefficients in front of $(\varphi^{+}\varphi)^{2}/24$
and $\bar A_{0}^{2}\varphi^{+}\varphi /8$, respectively.)

At this stage we can make three comments on the result. Firstly, note
that identically $d=0$, there is no direct induced sixth order selfcoupling
for $A_0$. Secondly,
the numerical coefficients in front of the couplings are small, which
diminishes the impact of these corrections even in the
non-asymptotic coupling range. Finally,
one observes that the ground state $\varphi =\varphi_{0}, \bar A_{0}=0$
 might become unstable for large values of $\varphi_{0}$
in certain ranges of the couplings due to the
the sixth power correction to the potential. This is probably
compensated by higher powers
of the expansion, as has been observed also in other cases [7]. It does not
cause here any problem, since this piece of the potential turns out
not to be important for the range of $\varphi$-values of interest for
the phase transition. In the range of the couplings $\lambda << g^{2}$, where
the sixth power term plays observable role, its coefficient is positive.

In case of U(1) theory an expression parallel to (8) has been
published before in [13]. The result
including the dim 4 part of the potential is:
$$L_{{\rm kin}}^{U(1)}={1\over 4}\bar F_{ij}\bar F_{ij}+{1\over
2}|(\partial_{i}+i
\bar g\bar A_{i})\varphi |^{2}+{1\over 2}(\partial_{i}\bar A_{0})^{2},$$
$$U_{{\rm dim}\,4}^{U(1)}(\bar A_0,\varphi)
={1\over 2}m_{\phi}^{2}|\varphi|^{2}+{1\over
2}m_{D}^{2}\bar A_{0}^{2}+{T\hat\lambda\over 24}|\varphi|^{4}+{\bar
g^{4}\beta\over 24\pi^{4}}\bar A_{0}^{4}+{1\over 2}\bar g^{2}\bar A_{0}^{2}|
\varphi|^{2}$$
$$~~~~~~~~~~~~~~~+{\rm 3D~counter terms},\eqno(20)$$
where the fields and the couplings are related to their 4D
counterparts as in (17), while
$$m_{\phi}^{2}=\hat m^{2}+({2\bar\lambda\over 3}+3\bar g^{2}){T\over
12},~~~~m_{D}^{2}=\bar g^{2}({T\over 3}+{m^{2}\over
4\pi^{2}T}),$$
$$\hat m^{2}=m^{2}\bigl(1-{1\over 8\pi^{2}}({9g^{4}\over 2\lambda}
+{2\lambda\over 2}+2g^{2})
+{1\over 16\pi^{2}}
(\lambda\ln{\lambda v_{0}^{2}\over 3T^{2}}+2(3g^{2}+{\lambda\over 6})
\ln{3g^{2}v_{0}^{2}\over T^{2}})\bigr),$$
$$\hat\lambda =
\lambda -{9\over 16\pi^{2}}(3g^{4}+g^{2}\lambda+{5g^{2}\over 18})$$
$$~~~~~~~~~+
{3\over 8\pi^{2}}\bigl( 2g^{4}(\ln {g^{2}v_{0}^{2}\over T^{2}}-4\ln{2\sqrt{2}
g^{2}v_{0}^{2}\over T^{2}})+{\lambda^{2}\over 4}\ln{\lambda v_{0}^{2}
\over 3T^{2}}+(3g^{2}+{\lambda\over 6})^{2}\ln{3g^{2}v_{0}^{2}\over T^{2}}
\bigr).\eqno(21)$$

The list of dim 6 corrections in case of the Abelian Higgs model
is the following:
$$\delta m_{\phi}^{2}={m^{4}\zeta (3)\over 384T^{3}\pi^{4}}(-9\bar
g^{2}+\bar\lambda ),$$
$$\delta m_{D}^{2}=-{m^{4}\zeta (3)\bar g^{2}\over 32T^{3}\pi^{4}},$$
$$\delta\bar\lambda ={m^{2}\zeta (3)\over 32T^{3}\pi^{4}}(-9\bar
g^{4}-3\bar g^{2}\bar\lambda+{5\over 6}\bar\lambda^{2}),$$
$$\delta\bar g^{2}=-{m^{2}\zeta (3)\bar g^{2}\bar\lambda\over
48T^{3}\pi^{4}},$$
$$a={\zeta (3)\over 256\pi^{4}}(\bar g^{6}-{\bar g^{4}\bar\lambda\over
2}-{\bar g^{2}\bar\lambda^{2}\over 12}+{7\bar\lambda^{3}\over 162}),$$
$$b={\zeta (3)\bar g^{2}\over 64\pi^{4}}(-{\bar g^{4}\over
4}+{\bar\lambda\bar g^{2}\over 6}-{\bar\lambda^{2}\over 9}),$$
$$c=d=0.\eqno(22)$$

The 1-loop integration of the effective 3D model which takes into
account the corrections (19) or (22) will be compared to the solution
where the potential is truncated at {\it dim 4} level (e.g. eqs.(14) or
(20)). We recall, the latter approximation is equivalent to
the "daisy-summed" infrared
improved treatment of the full 4D thermal theory.
The corrections we are going to discuss are formally of higher order than the
${\cal O}(\lambda^{3/2})$ and ${\cal O}(g^{3})$ corrections, which are
produced in the 1-loop
solution of the effective model, but their actual importance
 can be assessed only
if the physical couplings and $T=0$ mass values of the W-boson and
of the Higgs particle are substituted. It is convenient to compute
the 1-loop effective Higgs-potential from the 3D theory in the 3D
analogue of the 't Hooft-Landau gauge. We are aware of the fact that
especially the higher dimensional pieces of the potential might display
some gauge dependence, which, however should not affect  physical
quantities determined in any fully fixed gauge.
Using the notations
introduced above the corrected form of the effective potential
can be written (almost) uniformly for the U(1) and SU(2) cases:
$$U_{3D,eff}^{1-loop}={1\over 2}(m_{\phi}^{2}+\delta
m_{\phi}^{2})\phi_{0}^{2}+{1\over 24}(\hat \lambda
+\delta\lambda)\phi_{0}^{4}+{a\over T^{2}}\phi_{0}^{6}$$
$$~~~~~~~~~~-{T\over 12\pi}(n_{W}[Q(g^{2}+\delta
g^{2})]^{3/2}\phi_{0}^{3}+n_{A_{0}}\bar m_{D}^{3}+n_{G}\bar
m_{G}^{3}+\bar m_{H}^{3}),\eqno(23)$$
with $n_{W}=6~(2), n_{A_{0}}=3~(1), n_{G}=3~(1)$ for SU(2) (U(1)), and
$$\bar m_{H}^{2}=m_{\phi}^{2}+\delta m_{\phi}^{2}+{1\over 2}(\hat\lambda
+\delta\lambda )\phi_{0}^{2}+30a\phi_{0}^{4},$$
$$\bar m_{G}^{2}=m_{\phi}^{2}+\delta m_{\phi}^{2}+{1\over 6}(\hat\lambda
+\delta\lambda )\phi_{0}^{2}+6a\phi_{0}^{4},$$
$$\bar m_{D}^{2}=m_{D}^{2}+\delta m_{D}^{2}+Q(g^{2}+\delta
g^{2})\phi_{0}^{2}+2b\phi_{0}^{4}.\eqno(24)$$
($Q=1/4~(1)$ for SU(2) (U(1)).) In eqs. (23) and (24) 4-dimensional
notations are reinstated.

The degeneracy temperature of the effective potential (23) has been
found in various points of the ($\lambda , g$)-plane.
Several physical quantities were evaluated for the characterization of
the strength of the phase transition. The order parameter discontinuity
was found by locating
the non-trivial degenerate minima of the potential at $T_{c}$, the relative
surface tension $\sigma /T_{c}^{3}$ was calculated from the formula
valid in the thin--wall approximation. Thermal masses at the transition
are computed from (24) and
$$m_{W}^{2}=Q(g^{2}+\delta g^{2})\phi_{c}^{2}.\eqno(25)$$

In Table 1 results are presented for G=SU(2) with fixed input data $(g=2/3,
v_{0}=241.8$ GeV) and varying $m_{H}(T=0)$. These quantities in turn
determine $\lambda , m_{W}(T=0)$ through the usual relations.
All physical quantities are
expressed in proportion to the $T=0$ Higgs condensate
$v_{0}$, if not stated otherwise.
The results of the first row for each Higgs mass value were obtained
from a version of the effective theory with only ${\cal O}(T^{2})$
mass-corrections and
unmodified $\lambda$. (This version could be interpreted also
as normalised at the common scale $\mu =T$, and assuming the unchanged
relationship of the $T$=0 masses and the renormalised couplings granted.)
In the second row the results represent the effect
induced by the Linde-type renormalisation of the $T$-independent part of the
potential (e.g. eq.(9)). In the third row the $\phi_{0}^{6}$-type
contributions to the potential
are included, finally ${\cal O}(m^{2}/T^{2})$ corrections to the masses and
$\lambda$ are taken into account in the entries of the fourth row. All
quantities point to the well-known conclusion that the first order
nature of the transition weakens with increasing $m_{H}(T=0)$.

Numerically important effects come when the logarithmic renormalisation
corrections become large.
For small ($\leq 35$GeV) Higgs masses they make the transition
of harder first order nature. They signal large deviation of the physically
relevant scale $T_c$ from the normalisation scale $v_0$.
As a consequence also the effect of the higher dimensional
$\phi^{6}$ potential piece becomes important as it is  illustrated
for $m_{H}=20$ GeV in Fig.2.
 The mass-corrections are everywhere negligible as one recognizes also from the
small $\phi$ region of the figure.

 Experimental data seem to exclude the region below 50 GeV
for the Higgs mass. Therefore, we conclude that all formally higher order
corrections (induced by the renormalisation conditions and/or
by taking into account higher dimensional interactions
and  mass corrections)
generated in the reduction procedure are indeed negligible according to the
perturbative discussion of the phase transition.

We conclude this section by presenting the evidence for the existence of
distinct mass scales within the effective static theory as it emerges from its
perturbative solution.

Since this search cannot be based on any {\it a priori} theoretical
guiding principle, it starts with thorough inspection of the thermal
mass values of the different
fluctuation modes found at $T_{c}$
perturbatively. Any suggestion for the
existence of well-separated scales should be checked for
selfconsistency. This means that the solution of a theory resulting
from any further reduction, based on the existence of a conjectured
mass hierarchy, should preserve that feature in its output.

In Fig. 3 the variation of the thermal masses of the 4 different
fluctuations are depicted as functions of $m_{H}(T=0)$ at $T=T_{c}$, as
found from (24-25). It is obvious,
that $m_{D}$ is separated from the actual lightest mode(s) everywhere.
On the other hand the screening masses of the magnetic gauge and of the
scalar modes show an interesting cross-over around 50 GeV. For Higgs
masses much below this value it is very plausible to conjecture the
existence of a description in terms of a scalar effective theory
[11].

With some empathy for $m_{H}(T=0) >~ 80~ {\rm GeV}$ the 1-loop perturbative
screening
mass values can be seen to indicate the relevance of a pure gauge theory, which
might faithfully account for the features of the phase transition. In
the second part of this study the crudest realisation of the effective
pure gauge theory will be presented and its perturbative solution be discussed.
\bigskip
{\bf 3. Effective Gauge Field -- Higgs Field and
Pure Gauge Field Theories of the Electroweak Phase Transition}
\bigskip
The order of integration over the $\bar A_0^a$ and $\varphi$-fields will
follow the hierarchy of their thermal masses established from Fig.3 at
$T=T_{c}$. 
Thus first $\bar A_0^a$, then $\varphi$, is integrated over.
Both integrations can be performed without fixing a gauge.
However, the $\varphi =\varphi_{0}=$ constant and the slowly varying
$\bar A_{i}^{a}(x)$ background fields both break gauge invariance, therefore
the result can be interpreted physically only when the gauge is fully fixed.
This will be done in section 4.

The integrations will be performed in the Gaussian approximation.
Although the use of more sophisticated approaches might prove unavoidable
in later investigations, the intuition based on experience in
statistical physics tells, that multiple application of relatively
inaccurate renormalisation group steps might lead to quite accurate
effective dynamics of the lowest modes.

The quadratic form of the part of the action, relevant for the Gaussian
$\bar A_{0}^{a}$-integration reads, after Fourier-transforming it, as
$$S_{A_{0}}^{(2)}={1\over 2}\sum_{k}\bar A_{0}^{a}\bigl
[(k^{2}+m_{D}^{2}+{1\over 4}\bar g^{2}\varphi^{2}+\bar
g^{2}\bar (A_{i}^{c})^{2})\delta^{ab}-2i\bar g\epsilon^{acb}k_{i}
\bar A_{i}^{c}-\bar
g^{2}\bar A_{i}^{a}\bar A_{i}^{b}\bigr ]\bar A_{0}^{b}.\eqno(26)
$$
No appealing representation could be found for the determinant of the
$3\times 3$ fluctuation matrix, but the first terms
of its expansion with respect to $\bar A_{i}$ are as follows:
$$
\Delta_{A_0}U[\varphi ,\bar A_i]=-{1\over 4\pi}(m_{D}^{2}+{1\over 4}\bar
g^{2}\varphi^{2})^{3/2}+{\bar g^{4}\over 96\pi}{(\bar A_{i}\times
\bar A_{j})^{2}\over (m_{D}^{2}+{1\over 4}\bar g^{2}\varphi^{2})^{1/2}}+
{\cal O}(\bar A^{6}).\eqno(27)
$$
No mass term is generated in this step for $\bar A$. The quartic piece is
interpreted as the correction to the pure gauge kinetic action (no need
for explicit derivative expansion!):
$$
{1\over 4}\bar F_{ij}^{a}\bar F_{ij}^{a}\rightarrow {1\over
4\mu}\bar F_{ij}^{a}\bar F_{ij}^{a},$$
$${1\over \mu}=1+{g^{2}T\over 24\pi(m_{D}^{2}+{1\over
4}g^{2}\varphi^{2})^{1/2}}.\eqno(28)$$

The effective 3D gauge field-Higgs-model obtained after integrating over
$\bar A_{0}$ is then defined by the following action:
$$
S[\bar A_{i},\varphi]=\int d^3x \bigg[
{1\over 4\mu}\bar F_{ij}^{a}\bar F_{ij}^{a}
+{1\over 2}(D_{i}\varphi )^{+}(D_{i}\varphi )+U(\varphi ,\bar A_i)\biggr],
$$
$$
U(\bar A_i,\varphi )={1\over 2}m_{\phi}^{2}\varphi^{+}\varphi+
{\hat\lambda\over 24}(\varphi^{+}\varphi )^{2}-{1\over 4\pi}(m_{D}^{2}+
{1\over 4}\bar g^{2}\varphi^{2})^{3/2},\eqno(29)
$$
with $\mu$ given by eq.(28).

Here we derived the effective action by an explicit integration over the
$A_0^a$ field. Another way of proceeding would be to start from the static
effective action in eqs.~(14)-(17) and to evaluate two Feynman diagrams: the
$A_0^a$ loop contribution to the $A_i^a$ and $\varphi$ propagators.
The $k^2$ term of the former gives the $1/\mu$ in eq.~(28) (for $\varphi
\ll m_D$) and the latter gives the correction $m_{\phi}^2\to m_{\phi}^2
-3\bar g^2 m_D/(16\pi)$ obtained also from eq.~(29) for $\varphi
\ll m_D$.

The expansion around $\varphi_{0}$ up to quadratic terms in an
$\bar A_{i}$ background gives the following effective masses for the
Higgs $(H)$ and the (pseudo)-Goldstone ($G$) modes:
$$\tilde m_{H}^{2}\equiv m_{H}^{2}+{1\over 4}\bar
g^{2}\bar A^{2},
{}~~~\tilde m_{G}^{2}\equiv m_{G}^{2}+{1\over 4}\bar g^{2}\bar A^{2},
$$
$$m_{H}^{2}=m_{\phi}^{2}+{\hat\lambda T\over
2}\varphi_{0}^{2}-{3\bar g^{2}\over 16\pi}(m_{D}^{2}+{1\over 4}\bar
g^{2}\varphi_{0}^{2})^{1/2},$$
$$
m_{G}^{2}=m_{\phi}^{2}+{\hat \lambda T\over
6}\varphi_{0}^{2}-{3\bar g^{2}\over 16\pi}[(m_{D}^{2}+{1\over 4}\bar
g^{2}\varphi_{0}^{2})^{1/2}+{1\over 4}\bar
g^{2}\varphi_{0}^{2}(m_{D}^{2}+{1\over 4}\bar
g^{2}\varphi_{0}^{2})^{-1/2}].
\eqno(30)$$
It is remarkable that the corrections to the effective masses arising
from the expansion of (27) reproduce correctly the $A_{0}$-loop
contribution to the scalar selfenergies [3,4].

Integrating finally over $\varphi$ one obtains the following
fluctuation determinant: 
$$\Delta_{\varphi}U(\bar A_i)={1\over 2}\int {d^{3}k\over (2\pi)^{3}}\bigl\{
\log\bigl [(k^{2}+\tilde m_{G}^{2})(k^{2}+\tilde m_{H}^{2})-\bar
g^{2}(k_{i}\bar A_{i})^{2}\bigr ]+$$
$$~~~~~~~~~~~~~~+\log\bigl [(k^{2}+\tilde m_{H}^{2})^{2}-\bar
g^{2}(k_{i}\bar A_{i})^{2}\bigr ]\bigr\},\eqno(31)$$
which after the expansion into powers of $\bar A_{i}$ yields the following
contribution to the pure gauge action:
$$\Delta_{\varphi}
U(\bar A_i)=-{1\over 12\pi}(m_{H}^{3}+3m_{G}^{3})+{1\over
2}\bar A^{2}{\bar g^{2}\over 48\pi}{(m_{G}-m_{H})^{2}\over
m_{G}+m_{H}}$$
$$~~~~~~~~~~+{\bar g^{4}\over 48\pi}(\bar A_{i}\times \bar A_{j})^{2}
\bigl [{1\over 5(m_{G}+m_{H})}+{m_{G}m_{H}\over
5(m_{G}+m_{H})^{3}}+{1\over 8m_{G}}\bigr ]$$
$$~~~~~~~~~~-{\bar g^{4}\bar A^{4}(m_{G}-m_{H})^{2}\over
2560\pi m_{G}m_{H}(m_{G}+m_{H})^{3}}\bigl
[5(m_{G}^{2}+m_{H}^{2})+6m_{G}m_{H}\bigr
]+{\cal O}(\bar A^{6}).\eqno(32)$$
The first term on the right hand side of (32) is the usual 1-loop
contribution of the scalar fluctuations to the effective
$\phi_{0}$-potential within the 3D theory. The second and the fourth
terms (and probably many of the higher order terms too) violate gauge
invariance. These terms are proportional to $(m_{G}-m_{H})^{2}$,
what is clear manifestation of the fact that they are induced by the
symmetry breaking value
of the $\phi$-field. It is remarkable again that the second term
reproduces the contribution of the scalar loops to the polarisation
tensor of the magnetic gauge fluctuations [3,4].

It seems worthwhile to comment on the stability of the ground
state of eq. (32) with respect to $A^{2}$-fluctuations. In a
fluctuating $A$-background it is more appropriate to reexpress (32) in
terms of the field dependent masses $\tilde m_{G},\tilde m_{H}$,
defined in (30). This replacement to ${\cal O}(A^{4})$ affects directly only
the coefficient of $A^{2}$ in (32). Then one recognizes that the $\sim
A^{4}$ terms absorbed in this way from the higher power terms turn the
sign of the ${\cal O}(A^{4})$ term in the last line of (32) into positive.
Therefore (32) actually does not signal any vacuum instability in the
direction of the vector potential fluctuations. The same argument holds
also for the U(1) case in view of the remarks at the end of this
section.

{}From the third term of (32) one finds the contribution of the scalar
fields to the magnetic susceptibility of the gauge medium. This term is
proportional to $T/m_{G}$, therefore it requires careful
monitoring, in order to avoid the breakdown of the proposed
approximation, at least not for this specific reason.

For the discussion of the next section it is somewhat more convenient to
work with the usual form of the gauge kinetic term. This is achieved by
a finite rescaling of the vector potential, compensated in the potential
terms by a screened gauge coupling:
$$\tilde g^{2}={g^{2}\over 1+{g^{2}T\over 48\pi}\bigl ({1\over
2m_{G}}+{4\over 5(m_{G}+m_{H})}+{4m_{G}m_{H}\over
5(m_{G}+m_{H})^{3}}+{2\over (m_{D}^{2}+{1\over
4}g^{2}\phi_{0}^{2})^{1/2}}\bigr )}.\eqno(33)$$
In terms of the rescaled quantities the action of the effective pure
gauge action has the following form: 
$$S[\bar A_i]=\int d^{3}x\bigl\{{1\over
4}\bar F_{ij}^{a}\bar F_{ij}^{a}+{\tilde g^{2}T\over 2}\bar A_{i}^{2}({1\over
4}\varphi_{0}^{2}+{(m_{G}-m_{H})^{2}\over 48\pi
(m_{G}+m_{H})})$$
$$-{\tilde g^{4}T^{2}(\bar A^{2})^{2}(m_{G}-m_{H})^{2}\over
2560\pi
m_{G}m_{H}(m_{G}+m_{H})^{3}}[5(m_{G}^{2}+m_{H}^{2})
+6m_{G}m_{H}]+{\cal O}(\bar A^{6})+{\rm gauge~fixing~terms}\bigr \}.\eqno(34)$$
The effective theory in the broken phase differs from the 3D QCD
essentially because of the presence of $\varphi_{0}$-induced terms. It is
an interesting question to devise appropriate non-perturbative methods
for the study of the gauge dynamics in this phase.

Finally for completeness we quote the formulae resulting from a similar
procedure applied to the Abelian Higgs model. Since the additional
Higgs fields for the non-Abelian case actually contribute only to the
magnetic susceptibility (cf. the contribution of the second term
of (31)),
the effective pure gauge action is of the same form after a replacement
$g/2\rightarrow g$ is made and the appropriate screened coupling and
thermal masses are used in it:
$$\tilde g^{2}={g^{2}\over 1+{g^{2}T\over 15\pi}({1\over
m_{G}+m_{H}}+{m_{G}m_{H}\over
(m_{G}+m_{H})^{3}})},\eqno(35)$$
$$m_{H}^{2}=m_{\phi}^{2}+{\hat\lambda\over 2}\phi_{0}^{2}-{g^{2}T\over
4\pi}\bigl
[(g^{2}\phi_{0}^{2}+m_{D}^{2})^{1/2}+g^{2}\phi_{0}^{2}(g^{2}\phi_{0}^{2}
+m_{D}^{2})^{-1/2}\bigr ],$$
$$m_{G}^{2}=m_{\phi}^{2}+{\hat\lambda\over 6}\phi_{0}^{2}-{g^{2}T\over
4\pi}(g^{2}\phi_{0}^{2}+m_{D}^{2})^{1/2},\eqno(36)$$
(the formulae for $m_{D}^{2}, m_{\phi}^{2}$ are given in (21)).

The essential characteristics of the perturbative solution of the SU(2)
and U(1) cases are expected very similar, therefore in the short section 4 we
restrict our study to the non-Abelian case.
\bigskip
{\bf 4. Perturbative Discussion of the Effective Pure Gauge Theory}
\bigskip
In this section the 1-loop integration of the effective pure gauge
action is discussed in the Landau gauge. It is worthless to say, that
conclusions drawn from this approximate solution are of very restricted
conceptual interest.

In the Landau gauge the six transverse degrees of gauge freedom
fluctuate with the thermal mass
$$m_{W}^{2}={1\over 4}\tilde g^{2}\phi_{0}^{2}+{\tilde g^{2}T\over
48\pi}{(m_{H}-m_{G})^{2}\over m_{H}+m_{G}}.\eqno(37)$$
The fluctuations of the spatial longitudinal and the ghost degrees of
freedom are infinitely suppressed. At 1-loop level the higher power
$\bar A$-terms do not play any role.

The effective Higgs potential is obtained by adding to the gauge
contribution the $\bar A$-independent pieces from eqs. (27) and (32), and
also from the potential part of (14). The result is
$$U_{eff}(\phi_{0})={1\over 2}m_{\phi}^{2}\phi_{0}^{2}+{\hat\lambda\over
24}\phi_{0}^{4}-{T\over 12\pi}[6m_{W}^{3}+3(m_{D}^{2}+{1\over
4}g^{2}\phi_{0}^{2})^{3/2}+3m_{G}^{3}+m_{H}^{3}].\eqno(38)$$

On the basis of Fig.3 the proposed effective description is expected to
work for higher $m_{H}$ values. Therefore in Table 2 we compare physical
quantities characterising the phase transition for $m_{H}=80, 100, 120$
GeV as given by the potential (38) with the results from
the conventional perturbative treatment of the full 3D effective theory.
The parameters $m_{W}(T=0)=80.6$ GeV, $g=2/3$ are kept fixed close to their
experimental values.

First of all we can state that by the observed values of the screened
coupling the approximation is not endangered in this $m_{H}(T=0)$ region by
$T/m_{G}$ growing too large. The general characterisitics of all
the physical quantities hints to stronger discontinuities from the
effective gauge theory relative to the usual perturbative treatment. The
decrease in $T_{c}$ is 2-3\%, the increase of $\phi_{c}/v_{0}$ is 5-10\%,
and that of $\sigma /T_{c}^{3}$ is $\approx 50$\%.

On the other hand the gap between the scalar and the vector screening
masses closes considerably, at least in the present {\it perturbative}
solution of the effective pure gauge theory (Fig. 4). This fact can be traced
mainly to the negative mass contribution of the $\bar A_{0}$-loop to the scalar
masses (eq.(30)). Similar phenomenon led the authors of  Ref.[3] to
restrict the validity of the improved 1-loop perturbative treatment to
$m_{H}(T=0)< 100~{\rm GeV}$.
It could be that the opening of the gap is just shifted
to larger Higgs masses.

Finally, we should compare our findings with the available results of
non-perturbative investigations of the reduced 3D model [8,9].
For this purpose their data obtained at
$m_{H}=80~GeV$ seem to be the most relevant: $T_{c}=160~{\rm GeV},
\phi_{c}/T_{c}=0.67$. Both differ considerably from the perturbative
results, independently of whatever version of that latter is being
considered. In the effective pure
gauge model (with the renormalisation conditions (9))
 one has $T_{c}=184.5~{\rm GeV}, \phi_{c}/T_{c}=0.3$, which still
is shifted slightly towards the MC results from the output of the
usual perturbative treatment.

This circumstance underlines once again the interest of a
non-perturbative study of the effective pure gauge model.
Within the perturbative approach, higher loop corrections were
claimed to provide essential modification of the phase transition
characteristics [14]. Two-loop calculations relevant to the
analysis of the 3D effective models are in progress [15,16].
\bigskip
{\bf 5. Conclusions}
\bigskip
The interest in the three dimensional reduced effective theory of the
finite temperature electroweak interactions is obvious even from the great
simplification occuring in its non-perturbative study versus the simulation
of the full four-dimensional formulation [8,9]. Also attempts to exploit
 the extensive experience with critical phenomena in 3 dimensions need
a most faithful mapping onto a formally $T=0$ effective 3D theory [17].

In this paper we have studied a hierarchy of effective field theories
for hot electroweak matter. The hierarchy is based on integrating
subsequently over fields following the hierarchy of their masses.

The first stage was the standard dimensional
reduction [6]  in which nonstatic bosonic modes with temporal Fourier
index $n\not=0$ and with an effective mass of $2\pi nT$ were integrated
over. This leads to the effective action $S[A_i^a,A_0^a,\varphi]$ given
in eqs.~(14)-(17). We carried out the integration in a new way, which
permitted us to relate the 4D and 3D couplings and to study the magnitudes
of higher dimensional operators multiplied by inverse powers of $T$
(for dimensional reasons). We have restricted the range of importance
of the ${\cal O}(\log(m^2/T^2))$ and ${\cal O}(m^{2}/T^2)$ terms of the
effective action to well below the relevant experimental range of the mass of
the
Higgs particle ($m_H\leq 35 {\rm GeV}$). The line of thought presented here
can be followed without any change also for the more complete version of the
model, including the $U(1)$-component of the gauge field.

Subsequent stages were based on the hierarchy of masses exhibited in
Fig.~3. First, due to the large Debye mass $m_D$, the $A_0$ field is
integrated over, leading to the $S[A_i,\phi]$ given in eq.(29). The order
of further integrations is not so obvious. For Higgs masses above the
present lower limit it seems more accurate to integrate over the Higgs
field, leading to the effective action $S[A_i]$ given in eq.~(34). For smaller
Higgs masses integration over $A_i$ leads to an $S[\phi]$ studied in
[11]. Tendency for the hardening of the gauge-field driven phase transition
could have been observed when the final effective pure gauge theory has been
solved perturbatively for $m_H(T=0)\geq 80$GeV.

These rather reassuring results urge for the continuation of our investigation
in different directions. First of all, it would be very important to have
a systematic study of the gauge dependence of the reduced action, and
be able at the same time to argue for the gauge independence of the
phase transition characteristics on the basis of explicit calculations. The
second question of interest refers to the importance of higher loop effects
both in the reduction and in the solution of the effective theory. Thirdly,
at least the integration over the screened static $A_0$ field should be
possible to perform with an accuracy equal or better than that of the
non-static modes and its indirect effects on the effective Gauge-Higgs
dynamics be studied. Possible application of some sort
of low-$k$ cut-off integration technique [18] might prove useful in this
respect.


{\bf Acknowledgements} 
One of the authors (AP) would like to express his gratitude
to the Inst. of Theoretical
Physics of Univ. Bielefeld and of Univ. of Helsinki and to the Niels Bohr
Institute for the hospitality extended to him during different stages of this
calculation and to thank J. Ambjorn, P. Elmfors, N.-K. Nielsen,
A. Rebhan and P. Olesen for valuable discussions on the subject of this paper.
KK is deeply grateful to M. Shaposhnikov for discussions and advice.
\bigskip
{\centerline{\bf Appendix}
\bigskip
For the simple case of $A_{0}=0$ the matrices $K_{i}$ of eq.(6) take
block-diagonal form further facilitating the evaluation of their determinants:
$$\det K_{1}=\bigl( \det \left(\matrix{
\omega_{n}^{2}+M_{W}^{2}&{ig\over 2}k\Phi\cr
{-ig\over 2}k\Phi&K^{2}+M_{G}^{2}\cr}\right)
\bigr)^{2},\eqno(A.1)$$\
$$\det K_{2}=(K^{2}+M_{H}^{2})\det\left(\matrix{
\omega_{n}^{2}+M_{W}^{2}&{ig\over 2}k\Phi\cr
{-ig\over 2}k\Phi&K^{2}+M_{G}^{2}\cr}\right),\eqno(A.2)$$
$$\det K_{3}=(K^{2}+M_{W}^{2})^{2},~~~~\det K_4 =K^{2}+M_{W}^{2}\eqno(A.3)$$
($M_{W}^{2}=g^{2}\Phi^{2}/4,~M_{H}^{2}=m^{2}+\lambda\Phi^{2}/2,
{}~M_{G}^{2}=m^{2}+\lambda\Phi^{2}/6)$.

The 1-loop correction to the effective potential in the thermal static gauge
is written as
$$\Delta U=\int_{k}^{'}\{{3\over 2}\ln((K^{2}+M_{G}^{2})(\omega_{n}^{2}
+M_{W}^{2})-k^{2}M_{W}^{2})+{1\over 2}\ln (K^{2}+M_{H}^{2})+
3\ln (K^{2}+M_{W}^{2})\},\eqno(A.4)$$
where the meaning of the symbol $\int_{k}^{'}=T\sum_{n\neq 0}\int
d^{3}k/(2\pi)^{3}$.

One is allowed to expand the logarithms in powers of $M_{i}^{2}$. We work
up to ${\cal O}(M_{i}^{4})$ allowing to find the 1-loop corrections to the
classical Higgs potential. The field independent leading terms can be omitted
and the "primed" integrals of (A.4) are performed with help of the following
relations:
$$
\int_{k}^{'}{1\over K^{2}}={\Lambda^{2}\over 8\pi^{2}}-
{\Lambda T\over 2\pi^{2}}+{T^{2}\over 12},$$
$$\int_{k}^{'}{1\over K^{4}}={1\over 8\pi^{2}}+{D_{0}\over 8\pi^{2}},$$
$$\int_{k}^{'}{1\over \omega_{n}^{2}K^{2}}={\Lambda\over 24\pi^{2}T}-
{D_{0}\over 4\pi^{2}},\eqno(A.5)$$
with
$$D_{0}=\sum_{n\neq 0}\int_{0}^{\Lambda\beta}dx{2\over x^{2}+(2\pi n)^{2}}\sim
\ln{\Lambda\over T}.\eqno(A.6)$$
These steps lead to the detailed formula
$$\Delta U=({9\over 2}M_{W}^{2}+{3\over 2}M_{G}^{2}+{1\over 2}M_{H}^{2})
({\Lambda^{2}\over 8\pi^{2}}-{\lambda T\over 2\pi^{2}}+{T^{2}\over 12})
$$
$$~~~~~~~~+{1-D_{0}\over 16\pi^{2}}[{3\over 2}(M_{W}^{2}+M_{G}^{2})^{2}+
{1\over 2}
M_{H}^{4}+3M_{W}^{4}]-{3D_{0}\over 8\pi^{2}}M_{W}^{2}M_{G}^{2}$$
$$=\Phi_{0}^{2}[({9g^{2}\over 64\pi^{2}}+{\lambda\over 16\pi^{2}})\Lambda^{2}
-({9g^{2}\over 8}+{\lambda\over 2}){\Lambda T\over 2\pi^{2}}+
({9g^{2}\over 8}+{\lambda\over 2}){T^{2}\over 12}+
m^{2}({\lambda\over 16\pi^{2}}+
{3g^{2}\over 64\pi^{2}})$$
$$~~~~~~~~~-D_{0}m^{2}({\lambda\over 16\pi^{2}}+{9g^{2}
\over 64\pi^{2}})]$$
$$+\Phi_{0}^{4}[{\lambda^{2}\over 96\pi^{2}}+{9g^{4}\over 512\pi^{2}}+{
\lambda g^{2}\over 128\pi^{2}}-D_{0}({\lambda^{2}\over 96\pi^{2}}+
{9g^{4}\over 512\pi^{2}}+{3\lambda g^{2}\over 128\pi^{2}})]+
\Phi {\rm indep. terms}\eqno(A.7)$$
The logarithmic divergences originate from terms proportional to $D_{0}$.
\bigskip
{\bf Table Captions}
\bigskip
\item{Table 1} Phase transition characteristics from
various truncations of the effective 3D theory
a -- {\it dim 4} terms with $\mu =T$ renormalisation scale,
b -- {\it dim 4} terms with Linde-type renormalisation conditions,
c -- {\it dim 6} terms with Linde type renormalisation conditions,
d -- {\it dim 6} terms plus ${\cal O}(m^{2}/T^{2})$ mass and $\lambda$
corrections with Linde-type renormalisation conditions
\item{Table 2} Phase transition characteristics from the
3D pure gauge theory with $\mu =T$ renormalisation scale (b), and
with Linde-type renormalisation condition (c) compared with case
(a) of the previous table
\bigskip
{\bf Figure Captions}
\bigskip
\item{Fig. 1} Contour lines of the relative variation of a)
$(\hat m^{2}-m^{2})/m^{2}$, b)  $(\hat\lambda -\lambda)/\lambda$
covering the $m_H - T$ range relevant to the phase transition.
The various lines correspond to the 3\%,5\%,7\% levels, respectively.

\item{Fig. 2} Comparison of the effective potentials for the U(1) Higgs
model with {\it dim 4} and {\it dim 6} truncation
for $m_{H}(T=0)=20$GeV at the transition temperature of the {\it dim 6}
truncated model

\item{Fig. 3} Thermal (screening) masses of different fluctuations at
$T=T_{c}$ as functions  of $m_{H}$ in the SU(2) Higgs model
($v_{0}=241.8 {\rm GeV}, g=2/3$) calculated perturbatively form the {\it dim
4} truncated effective theory (Linde-type renormalisation conditions)

\item{Fig. 4} Thermal (screening) masses of different fluctuations at
$T=T_{c}$ as functions  of $m_{H}$ in the SU(2) Higgs model
($v_{0}=241.8 {\rm GeV}, g=2/3$) as calculated perturbatively from the
effective pure gauge theory.
\bigskip
{\bf References}
\item{[1]} G.W. Anderson and L.J. Hall, Phys. Rev. {\bf D45} (1992) 2685
\item{[2]} M.E. Carrington Phys. Rev. {\bf D45} (1992) 2933
\item{[3]} W. Buchm\"uller, Z. Fodor, T. Helbig and D. Walliser, DESY-93-021
\item{[4]} J.R. Espinosa, M. Quir\'os and F. Zwirner, Phys. Lett. {\bf B314}
(1993) 206
\item{[5]} M. Gleiser and E.W. Kolb, Phys. Rev. {\bf D48} (1993) 1560
\item{[6]} T. Appelquist and R. Pisarski, Phys. Rev. {\bf D23} (1981) 2305
\item{[7]} N.P. Landsman, Nucl. Phys. {\bf B322} (1989) 498
\item{[8]} K. Kajantie, K. Rummukainen and M. Shaposhnikov, Nucl. Phys.
{\bf B407} (1993) 356
\item{[9]} K. Kajantie, Talk at the 17th Johns Hopkins Workshop on Theoretical
Physics, 1993 July, Budapest (to appear in the Proceedings)
\item{[10]} A.D. Linde, Repts on Progress in Phys. {\bf 42} (1979) 389
\item{[11]} F. Karsch, T. Neuhaus and A. Patk\'os (Talk by T. Neuhaus at the
Lattice '93 Conference, Dallas 1993, October, to appear in the Proceedings)
\item{[12]} A.I. Bochkarev and M. Shaposhnikov, Mod. Phys. Lett. {\bf 2A}
(1987) 417
\item{[13]} A. Jakov\'ac and A. Patk\'os, Z. f. Physik {\bf C60} (1993) 361
\item{[14]} P. Arnold and O. Espinosa, Phys. Rev. {\bf D47} (1993) 3546
\item{[15]} K. Farakos, K. Kajantie, K. Rummukainen and M. Shaposhnikov,
CERN Preprint CERN-TH.6973/93
\item{[16]} A. Jakov\'ac and A. Patk\'os (work in progress)
\item{[17]} P. Arnold and L. Yaffe, Univ. of Washington preprint,
1993. November
\item{[18]} N. Tetradis and C. Wetterich, Nucl. Phys. {\bf B398} (1993) 659
\end